\documentclass{appolb}
\usepackage{epsfig}

\begin{document}
\title{Predictions from Microscopic Models on Particle Correlations at 
RHIC}
%
\author{Sven Soff
\address{Institut f\"ur Theoretische Physik, Goethe-Universit\"at, 60054 
Frankfurt am Main, Germany}
}
\maketitle
\vspace*{-1cm}  
\section{Introduction}
This contribution reviews the recent developments on microscopic 
transport calculations for    
two-particle correlations at low relative momenta in ultrarelativistic 
heavy ion collisions at RHIC. 
We will start with the discussion of a simple hadronic  
rescattering model, will then continue with the predictions of 
a hadron+string model (RQMD), and subsequently present the features 
of a combined hydrodynamical and microscopic hadron+string model 
(Hydro+UrQMD). We will then address the impact of partonic 
(elastic) scatterings (MPC) and also discuss the correlation 
results for a combined parton+hadron model (AMPT). 

For two-particle correlations of bosons at low relative momenta,  
Bose-Einstein correlations, i.e., the symmetrization of the 
two-particle wave functions, are usually the desired effect. 
One is interested in the space-time extent of the particle-emitting source. 
However, the situation is complicated by the strong time dependence of 
the dynamical multi-particle system. Thus, two-particle interferometry 
is not only sensitive to some geometric size parameters but also to 
the dynamics (for example, collective flow) and many other 
features of the source as the passing through a phase transition 
\cite{pratt86,Pratt:1990zq,Wiedemann:1999qn}. 
Indeed, it was predicted that, in case of a first-order phase transition,  
the correlation radii should be anomalously large due to the 
large latent heat that needs to be converted into the hadronic phase 
while the number of degrees of freedom 
is reduced simultaneously which in turn should 
lead to large hadronization times. 
A characteristic measure, being sensitive to the emission duration 
of the source, was thought to be the ratio of two correlation lengths 
$R_{\rm out}/R_{\rm side}$. 
The ratio was originally predicted to grow to large values like 5 or 10, reflecting the rather large 
emission times, for ultrarelativistic 
heavy ion collisions (for example Au+Au at RHIC, $\sqrt{s}_{nn}=200 \,$GeV), 
that is, for conditions where a deconfined phase should 
be produced.  
However, experimental results from RHIC \cite{STARpreprint,Johnson:2001zi} 
show ratios close to unity, even for 
various transverse momenta. One might conclude from this that 
hadronization times are short, possibly too short to support 
a first-order phase transition scenario. Indeed, there are indications 
(even though not settled yet) from  
lattice QCD that at RHIC (high T, low $\mu$) 
a cross-over transition might take place instead. 
For a thorough theoretical understanding of the underlying dynamics 
and the corresponding correlations one needs transport theory. 
Hydrodynamics has the advantage of its simple conceptual idea and 
the possibility to study explicitly the impact of different 
equations of state. However, two-particle interferometry at low relative 
momenta is also very sensitive to the dynamics close to the decoupling 
from the system (at the {\it freeze-out}). Microscopic transport theory 
allows one to calculate explicitly the freeze-out without relying on particular 
prescriptions. This provides one of many motivations to study the correlations 
in the framework of microscopic transport theory. 
Here, we systematically review the recent predictions from 
various microscopic transport models and discuss what can be learned from obvious 
discrepancies or seeming agreement. 

\section{Correlation functions and the $R_{\rm out}/R_{\rm side}$ ratio from microscopic freeze-out 
information}
The microscopic description provides discrete phase-space points for the last (strong) interactions. 
For Gaussian sources, the geometrical size parameters (correlation lengths) in the out-side-long 
coordinate system can be written as 
(see, for example, \cite{Wiedemann:1999qn})
\[
\hspace*{-1.2cm}
R_s^2=\langle \tilde{y}^2 \rangle,\hspace{1cm}
R_o^2=\langle(\tilde{x}-\beta_t\tilde{t})^2\rangle,\hspace{1cm}
R_l^2=\langle(\tilde{z}-\beta_l\tilde{t})^2\rangle,
\] 
where $\tilde{t},\tilde{x},\tilde{y},$ and $\tilde{z}$ are the space-time coordinates relative to 
the mean source centers $\tilde{x}_{\mu}=x_{\mu}-\langle \tilde{x}_{\mu} \rangle$. 
These expressions enlighten the mutual interplay of spatial and temporal components for the correlation radii and 
also enable their disentanglement. 
However, for a direct comparison to experimental data the explicit calculation of correlation functions 
\[
\hspace*{-1.2cm}C_2-1 \simeq \frac{\int d^4x S(x,{\bf K})
\int d^4y S(y,{\bf K})
\exp[2ik\cdot(x-y)]}
{|\int d^4x S(x,{\bf K})|^2}
\]
is necessary \cite{Hardtke:1999vf,kaonlett,Soff:2002pc,Lin:2002gc}. 
$S(x,K)$ represents the classical source function depending on position and momentum. 
The $C_2$'s are subsequently fitted to a Gaussian form of the correlator, 
\[
C_2=1+\lambda \exp(-q_o^2R_o^2 -q_s^2
R_s^2-q_l^2R_l^2)
\]
This procedure is the standard method how to extract the correlation radii. 
     
As a first example we report the findings obtained with a simple 
{\bf hadronic rescattering}  
model \cite{Humanic:2002iw}. Starting from an initial thermalized state at $\tau_{\rm had}=1\,$fm/c 
with a temperature $T=300\,$MeV and a width of a Gaussian rapidity distribution $\sigma_y=2.4$, 
hadrons are rescattering until they reach freeze-out. The parameters were chosen such that 
the final global observables are reproduced. Surprisingly, 
the results of this purely hadronic approach for the correlation radii show only small differences to the data. 
The ratio $R_{\rm o}/R_{\rm s}$ is only slightly larger than unity and the radii are roughly reproduced. 
This model relies on rather strong assumptions 
as the very fast hadronization or the existence of hadrons at large energy densities of several GeV/fm$^3$;  
hence, the interpretation of the obtained results is rather difficult although they contribute 
to the general picture of HBT correlations from microscopic models at RHIC.

The  {\bf RQMD} model, relying on (di)quark, hadron and string degrees of freedom,  
has also been applied to extract the correlation radii \cite{Hardtke:1999vf}. 
The fitted radii are again roughly reproduced (the calculated radii are typically $\sim 1\,$fm larger 
than the exp.\ data), the ratios are 1.2--1.4 for the calculations and closer to unity 
for the data (0.9--1.1) by possibly exhibiting an opposite $p_t$-dependence.
This model does not {\it explicitly} include a phase transition to quark-gluon matter. 
However, during the high density phase the effective degrees of freedoms are 
string excitations and constituent (di)quarks. 
The finite (default) formation times (until string fragmentation) may lead to a lack of pressure 
(that are reflected in elliptic flow values smaller than data). 
Still the obtained radii do show reasonable magnitudes indicating that the freeze-out dynamics 
may dominate the HBT radii. 
Using such a sophisticated microscopic model only for the later, more dilute stages 
of the system evolution and describing the high density phase by hydrodynamics has several 
advantages. One can explicitly study the dependencies on the equation of state by simultaneously 
following a realistic freeze-out. 
The {\bf UrQMD} model has been coupled at the hadronization hypersurface 
to an initial \mbox{\bf hydrodynamic scaling flow} 
for the early phase \cite{Soff:2002pc,soffbassdumi,kaonlett}.   
First-order phase transition scenarios with different values for the critical temperature $T_c$ ($\sim$ latent heat) 
have been studied. The fitted $R_s$ radii agree with data while $R_o$ (and $R_l$) are overpredicted ($\sim 20-30\%$).
Several studies have been performed to check possible improvements. 
In-medium modifications (of the $\rho$ meson), for example, improve the comparison (in particular for $R_l$) \cite{Soff:2002pc}
that correspond to effectively increased opacities. Another sensitivity, that to the transition temperature 
$T_{\rm switch}$ (hydro-micro) is shown in the figure. The comparison to data improves if this transition 
is performed later in the hadronic phase ($T_{\rm switch}=130\,$MeV) instead of the default 
transition temperature at hadronization ($T_{\rm switch}=T_c=160\,$MeV). 
Then, the non-ideal microscopic phase is shorter, reducing the radii. 

\begin{figure}[htp]
\vspace*{-.4cm}
\centerline{\hspace{.4cm}\hbox{\epsfig{figure=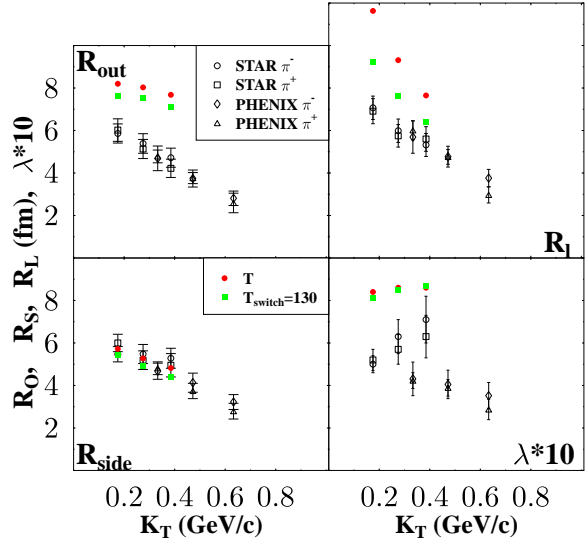,width=8cm}}}
\caption{HBT-radius parameters $R_{\rm out},\,R_{\rm side},\,
R_{\rm long},$ and intercept parameter $\lambda$ as a function 
of the transverse momentum $K_T$ as calculated from 
the phase space distribution of the 'QGP+hadronic rescattering' 
model (Hydro+UrQMD) 
for central nucleus nucleus collisions at RHIC compared to data from STAR and PHENIX. 
The transition temperature is varied from $T_{\rm sw}=T_c=160\,$MeV (full circles) to  $T_{\rm sw}=130\,$MeV 
(full squares).}
\end{figure}

Eventually, we discuss models that are (partially) based on 
partonic degrees of freedom. 
The {\bf MPC} models a classical gluon gas including elastic collisions 
\cite{Molnar:2002bz}. 
The main ingredient studied is the 
transport opacity $\chi\sim \sigma_{tr} \cdot dN_g(\tau_0)/d \eta$ 
being proportional to the 
transport cross section and gluon density. 
It has to be emphasized that no hadronic phase or resonances 
have been taken into account. Thus, 
the pure parton dynamics effects are studied without asking 
whether these dependencies survive 
a possible subsequent hadronic phase and/or resonance effects. 
Larger opacities naturally lead to later decoupling times. 
The calculated radii increase with opacity ($R_o,\,R_l$) but are still 
smaller than data (opposite to ideal hydro). 
$R_s$ seems to be unaffected by the early parton dynamics, i.e., 
it does not depend on $\chi$. 
While these trends are important for demonstrating the sensitivities 
of HBT to the early dynamics it has to be kept in mind that for 
a quantitative comparison to data hadrons/resonances have to be taken 
into account (beyond the simplified mapping gluon$\rightarrow \pi$). 

Finally, {\bf AMPT}, a combined model of initial (hard+soft) collisions, 
elastic parton scatterings, and a hadron cascade has been used 
to study the impact of different elastic parton cross sections and 
the so-called string fusion mechanism (the parton content of the 
soft strings is required to participate in the 
parton cascade) \cite{Lin:2002gc}. 
With this mechanism and $\sigma_{\rm part}\approx 10\,$mb      
the calculations come closest to the data, in particular to 
$R_o/R_s\approx 1$. It is interesting to note that within 
this model the $x_{out},t$-correlations are positive. Hence, they 
give a negative contribution to $R_o$ (see Eq.\ 1, mixed 
term of $R_{\rm out}$). 
This seems to be different, in particular, from the cross correlations 
in hydrodynamical models which are negative.
  
\section{Conclusions}
There has been enormous progress in the approaches to understand the 
wealth of correlation data. Several new transport models have been 
applied to the correlation analysis. 
The overall magnitude of the correlation radii is understood. 
The {\it HBT-puzzle}, that is, the detailed $p_t$-dependence of the
$R_{\rm o}/R_{\rm s}$ ratio, however, 
cannot be considered to be fully solved and needs further studies 
\cite{Hardtke:1999vf,kaonlett,Soff:2002pc,Lin:2002gc,
Humanic:2002iw,soffbassdumi,Molnar:2002bz,Zschiesche:2001dx}. 
The importance of the early stage opacity and parton cross 
sections has been demonstrated. Similarly, the late stage decoupling, 
opacities, in-medium effects play an important role. 
The space-time correlations ($x_{\rm out},t$) 
represent an important contribution that need to be checked independently, 
for example, by means of nonidentical particle pair correlations.

\end{document}